\documentclass[prb,twocolumn,showpacs]{revtex4}
\usepackage{graphicx,epsf}

\newcommand{\bq}{{\bf q}}
\newcommand{\bk}{{\bf k}}
\newcommand{\bp}{{\bf p}}
\newcommand{\br}{{\bf r}}
\newcommand{\nubar}{\bar{\nu}}
\newcommand{\rhobar}{\bar{\rho}}
\newcommand{\chibar}{\bar{\chi}}
\newcommand{\Hhat}{\hat{H}}

\begin{document}

\def\tende#1{\,\vtop{\ialign{##\crcr\rightarrowfill\crcr
\noalign{\kern-1pt\nointerlineskip}
\hskip3.pt${\scriptstyle #1}$\hskip3.pt\crcr}}\,}

\title{Competition between quantum-liquid and electron-solid phases in intermediate Landau levels}
\author{M.\ O.\ Goerbig$^{1,2}$, P.\ Lederer$^2$, and  
C.\ Morais\ Smith$^1$}

\affiliation{$^1$D\'epartement de Physique, Universit\'e de Fribourg, P\'erolles,  CH-1700 Fribourg, Switzerland.\\
$^2$Laboratoire de Physique des Solides, Bat.\,510, UPS (associ\'e au CNRS), F-91405 Orsay cedex, France.}

\begin{abstract}
On the basis of energy calculations we investigate the competition between quantum-liquid and electron-solid phases in the Landau levels $n=1,2$, and $3$ as a function of their partial filling factor $\nubar$. Whereas the quantum-liquid phases are stable only in the vicinity of quantized values of $\nubar=1/(2s+1)$, an electron solid in the form of a triangular lattice of clusters with a few number of electrons (bubble phase) is energetically favorable between these fillings. This alternation of electron-solid phases, which are insulating because they are pinned by the residual impurities in the sample, and quantum liquids displaying the fractional quantum Hall effect explains a recently observed reentrance of the integral quantum Hall effect in the Landau levels $n=1$ and $2$. Around half-filling of the last Landau level, a uni-directional charge density wave (stripe phase) has a lower energy than the bubble phase. 

\end{abstract}
\maketitle

\section{Introduction}

Two decades after the discovery of the integral and fractional quantum Hall effects (IQHE and FQHE), two-dimensional (2D) electron systems in a perpendicular magnetic field remain a field of research with unforeseen and surprising phenomena.\cite{perspectives} Experimental findings range from the observation of large anisotropies in the longitudinal magneto-resistance\cite{exp1} to an intriguing reentrant IQHE (RIQHE) at moderate magnetic fields. \cite{exp2,exp3} From a theoretical point of view, these systems are particularly interesting because they exhibit a large variety of quantum phases ranging from quantum liquids responsible for the FQHE to electron-solid phases such as the Wigner crystal and charge density waves (CDWs). 
These quantum phases arise because in the presence of a magnetic field the electrons' kinetic energy is quantized in equidistant energy levels, which are called Landau levels (LLs). The separation between these highly degenerate levels is $\hbar\omega_C$, where $\omega_C=eB/m$ is the cyclotron frequency for electrons with charge $e$ and band mass $m$. The level degeneracy is characterized by the LL filling factor $\nu=n_{el}/n_B$, where $n_{el}$ is the planar electronic density and $n_B=1/(2\pi l_B^2)=eB/h$ is the flux density, with the magnetic length $l_B=\sqrt{\hbar/eB}$. The flux density determines the density of states per LL. Furthermore, each LL is split into two spin branches with an energy gap $\Delta_Z$ because of the Zeeman effect. Correlation effects due to the Coulomb interaction between the electrons become important if $\nu\neq N$, with integral $N$, and if the characteristic Coulomb energy $e^2/\epsilon l_B$ is smaller than the level separations $\hbar\omega_C$ and $\Delta_Z$, which is the case at large magnetic fields. 

In the two lowest LLs, strong correlations give rise to quantum-liquid phases, which display the FQHE at $\nubar=p/(2ps\pm 1)$, where $p,s$ are integers and $\nubar=\nu-N$ is the filling factor of the last, partially filled spin branch of the $n$-th LL. $N$ denotes the number of completely filled levels, which may be treated as inert, and obeys $N=2n$ for a partially filled lower spin branch and $N=2n+1$ for the upper spin branch of the $n$-th LL. The FQHE is understood in terms of composite fermions (CFs),\cite{jain} which are formed to reduce the repulsive Coulomb interaction between the electrons. This reduction is due to a factor $\prod_{i<j}(z_i-z_j)^{2s}$ in the $N$-particle wave function, where $z_j=x_j+iy_j$ is the position of the $j$-th particle in the complex plane. Furthermore, these CFs experience a reduced coupling $(eB)^*=eB/(2ps\pm 1)$ to the external magnetic field, and thus fill CF-LLs with a filling factor $\nu^*=n_{el}h/(eB)^*$ as do electrons in the case of the IQHE. For $p=1$, the CF wave functions coincide with Laughlin's trial wave functions, which describe the FQHE at $\nu=1/(2s+1)$.\cite{laughlin} Both the IQHE and the FQHE are usually explained by single-particle localization due to underlying impurities:\cite{trugman} if the filling factor is slightly increased from $\nu=N$ ($\nu^*=p$), {\sl e.g.} by lowering the magnetic field, electrons (CFs) are promoted to the next (CF-) LL. They first populate the minima of the underlying impurity potential and are thus localized so that they do not contribute to the electrical transport. The longitudinal and the Hall resistances therefore remain at their original values. This gives rise to the plateaus in the Hall resistance at values $R_{xy}=h/e^2N$ for the IQHE ($R_{xy}=(h/e^2)(2ps\pm 1)/p$  for the FQHE) if plotted as a function of the magnetic field accompanied by zeros in the longitudinal magneto-resistance. 

However, the insulating behavior of electrons in the last LL may also be due to another effect than single-particle localization. Because of their mutual Coulomb repulsion, the electrons in the last LL have a tendency to form a Wigner crystal at low $\nubar$. At larger values of $\nubar$, Hartree-Fock calculations indicate that in higher LLs these electrons may form CDWs.\cite{FKS,moessner,fogler} There is since quite long experimental evidence for the formation of a Wigner crystal in the lowest LL below $\nu\sim1/5$.\cite{jiang} Recent microwave experiments revealed that the Wigner-crystal phase around $\nu=N$ persists in higher LLs\cite{chen1} and that a triangular CDW (bubble) ground state is formed around $\nubar=1/4$ and $3/4$ in the $n=2$-LL.\cite{lewis} Another experimental evidence for CDW formation in higher LLs is the huge anisotropy observed in the longitudinal magneto-resistance at half-filling for $n\geq2$. \cite{exp1} In this case, the bubbles merge to form a uni-directional CDW (stripe phase). Because the electronic transport is easy along the stripes (low-resistance direction) and strongly suppressed across them, this effect gives rise to the anisotropy. 

Recent experimental investigations by Eisenstein {\sl et al.} in both spin branches of the $n=1$-LL have revealed a new intriguing phenomenon:\cite{exp3} between the FQHE states at $\nubar=1/5,1/3$ and the even-denominator state\cite{willet} at $\nubar=1/2$ with a quantized $R_{xy}=h/e^2(N+\nubar)$, the Hall resistance jumps back to values $R_{xy}=h/e^2N$, corresponding to the neighboring plateau of the usual IQHE. An analogous reentrant IQHE (RIQHE) had been observed before by Cooper {\sl et al.} in $n=2$ around $\nubar=1/4$ and $3/4$.\cite{exp2} As expected from the observation of the FQHE, which indicates the importance of strong correlations in this range of the filling factor, such insulating phases of electrons in the last LL are unlikely to be due to simple localization of single particles. 

In this article, we present detailed energy calculations of the different quantum-liquid and electron-solid phases in $n=1,2$, and $3$, which allow us to identify the electron-solid phase in form of a triangular CDW as the origin of this insulating behavior. The CDW is pinned by the underlying impurities and therefore does not contribute to the electrical transport\cite{FL1}. This leads to the same dc response in transport measurements as observed in the IQHE regime. However, in the vicinity of $\nubar=1/(2s+1)$ the quantum-liquid phases have a lower energy than the electron solid, provided that $\nubar$ is lower than a critical filling $\nubar_c(n)\propto 1/(2n+1)$, given as a function of the LL index $n$.\cite{goerbig2} This gives rise to the observed alternation between the different phases with first-order quantum phase transitions between them. The phases in $n=1$ and $n=2$ have already been discussed in a previous Rapid Communication.\cite{goerbig3} Here, we provide more details of the calculations including stripe phases and results for $n=3$. We further evaluate how pinning by impurities will affect the energy values of the electron-solid phases and discuss possible experimental consequences of our theoretical investigations. 

In Sec.\,II, we present a theoretical model for the description of electrons restricted to the $n$-th LL and its basic properties. The Hartree-Fock solutions of this model for triangular and uni-directional CDWs are obtained in Sec.\,III. The energy of the quantum-liquid phases exactly at $\nubar=1/(2s+1)$ is calculated in Sec.\,IV with the help of sum rules imposed on Laughlin's trial wave functions.\cite{goerbig1} If the filling factor is moved away from precisely these values, quasiparticles/-holes are excited. Their energies are calculated analytically in Sec.\,V in the framework of the Hamiltonian theory proposed recently by Murthy and Shankar.\cite{MS} In Sec.\,VI, a comparison of the energies of the different phases and the relation to experimental results are given. Impurity effects are investigated in Sec.\,VII, and possible experiments, which could verify our results, are discussed in Sec.\,VIII. A brief summary may be found in Sec.\,IX.

\section{The model}

We adopt a model of spinless electrons and restrict their dynamics to the $n$-th LL. This model is valid when the characteristic Coulomb-interaction energy $e^2/\epsilon l_B$ is smaller than the LL separation $\hbar\omega_C$ and the Zeeman splitting $\Delta_Z$. We further require that the partial filling factor of the last level $\nubar=\nu-[\nu]$ is different from zero because at integral fillings, the only possible low-energy excitations are inter-LL excitations, which cost an energy of order $\hbar\omega_C$ or $\Delta_z$. In the absence of inter-LL excitations, the number of electrons $\bar{N}_{el}$ in the last LL is fixed by the partial filling factor $\bar{N}_{el}=\nubar A/2\pi l_B^2$, where $A$ is the total area of the sample. The kinetic energy is therefore an unimportant constant $E_{kin}=\bar{N}_{el}\hbar\omega_C(n+1/2)$, which may be omitted in the Hamiltonian. The model is thus given by
$$\Hhat=\frac{1}{2}\int d^2rd^2r'\psi_n^{\dagger}(\br)\psi_n(\br)v(\br-\br')\psi_n^{\dagger}(\br')\psi_n(\br'),$$
where $v(r)$ is the usual Coulomb potential and the fermion fields contain only components of the $n$-th LL
$$\psi_n(\br)=\sum_{y_0}\langle \br|n,y_0\rangle c_{n,y_0}.$$
The wave functions in the Landau gauge ${\bf A}=B(-y,0,0)$ are given by $\langle \br|n,y_0\rangle=L^{-1/2}\exp(iy_0x/l_B^2)\chi_n(y-y_0)$ with $\chi_n(y)=(\sqrt{\pi}2^nn!l_B)^{-1/2}H_n(y/l_B)\exp(-y^2/2l_B^2)$, and $c_{n,y_0}^{\dagger}$ creates an electron in the state $|n,y_0\rangle$. In terms of Fourier components of the density operator
$$\rho_n(\bq)=\int d^2r\,\psi_n^{\dagger}(\br)\psi_n(\br)e^{i\bq\cdot\br}=F_n(q)\rhobar(\bq),$$
the Hamiltonian can be rewritten in reciprocal space as
\begin{equation}
\label{equ001}
\Hhat=\frac{1}{2}\sum_{\bq}v_n(q)\rhobar(-\bq)\rhobar(\bq),
\end{equation}
where $\sum_{\bq}=A\int d^2q/(2\pi)^2$. 
The projected density operator reads
$$\rhobar(\bq)=\sum_{y_0}e^{-iq_yy_0}c_{n,y_0+q_xl_B^2/2}^{\dagger}c_{n,y_0-q_xl_B^2/2},$$
and $F_n(q)=L_n(q^2l_B^2/2)\exp(-q^2l_B^2/4)$, where $L_n(x)$ is a Laguerre polynomial. The form factor $F_n(q)$ is due to the overlap of the wave functions in the $n$-th LL and has been absorbed in an effective interaction potential
\begin{equation}
\label{equ002}
v_n(q)=v(q)[F_n(q)]^2, \qquad v(q)=2\pi e^2/\epsilon q.
\end{equation}
Inter-LL excitations may be accounted for in a wave-vector-dependent dielectric function $\epsilon(q)$.\cite{FKS,AG} This dependence will be neglected in the further discussion because it has only a minor effect on the physical properties of the system as shown by Fogler and Koulakov.\cite{fogler}

\begin{figure}
\epsfysize+5.0cm
\epsffile{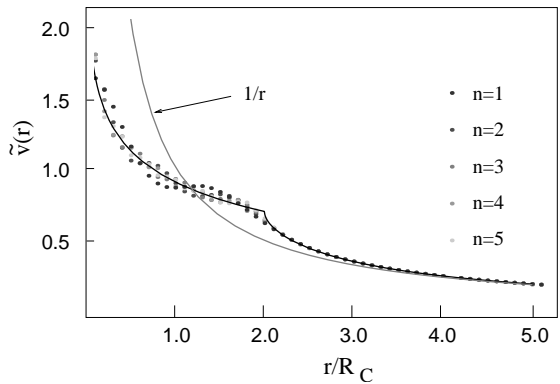}
\caption{Effective interaction potential $\tilde{v}(r)$ in real space in units of $e^2/\epsilon l_B$. Points correspond to $n=1,...,5$; gray line: pure Coulomb potential; black line: scaling form according to Eq.\,(\ref{equ0-1}).}
\label{fig00}
\end{figure}

With the help of the usual anti-commutation relations for the fermionic operators, $\{c_{n,y_0},c_{n',y_0'}^{\dagger}\}=\delta_{n,n'}\delta_{y_0,y_0'}$ and $\{c_{n,y_0},c_{n',y_0'}\}=\{c_{n,y_0}^{\dagger},c_{n',y_0'}^{\dagger}\}=0$, one finds that the projected density operators satisfy the algebra\cite{GMP}
\begin{equation}
\label{equ003}
[\rhobar(\bq),\rhobar(\bk)]=2i\sin\left(\frac{(\bq\times\bk)_zl_B^2}{2}\right)\rhobar(\bq+\bk).
\end{equation}
The physical origin of this algebra is the following: by projecting the density operators to the $n$-th LL, one averages over the rapid cyclotron motion so that $\rhobar(\bq)$ may be interpreted as a density operator of fermions, which are described only by their guiding-center coordinates ${\bf R}_i=\br_i-{\bf \hat{z}}\times{\bf p}_i/eB$ with $\br_i$ and ${\bf p}_i$ being position and momentum of the $i$-th particle, respectively. It is the non-commutativity of the components of the guiding-center coordinates $[X_k,Y_l]=-il_B^2\delta_{k,l}$, which is responsible for these commutation relations as well as for the fact that each electronic state occupies a minimal surface $\sigma=2\pi l_B^2=1/n_B$.

The Hamiltonian (\ref{equ001}) together with the commutation relations (\ref{equ003}) define the quantum mechanical model. The electrons interact via the effective interaction potential (\ref{equ002}), which contains the information of the LL of interest in the form factor $F_n(q)$. Deeper insight into the meaning of this interaction potential is obtained from a transformation back to real space, which is shown in Fig.\,\ref{fig00}. The effective interaction potential satisfies the scaling law $v_n(r)=\tilde{v}(r/R_C)/\sqrt{2n+1}$, where $R_C=l_B\sqrt{2n+1}$ is the cyclotron radius.\cite{goerbig2} With the help of $F_n(q)\simeq J_0(q\sqrt{2n+1})$, which becomes exact in the large-$n$ limit, one obtains
\begin{equation}
\label{equ0-1}
\tilde{v}(x)\simeq\frac{4e^2}{\pi\epsilon l_Bx}{\rm Re}\left[K\left(\frac{1-\sqrt{1-4/x^2}}{2}\right)\right]^2,
\end{equation}
where $J_0(x)$ is the zero-order Bessel function and $K(x)$ the complete elliptic integral of the first kind. Comparison with the exact form of the real-space interaction potential (Fig.\,\ref{fig00}) indicates that the scaling form (\ref{equ0-1}) is accurate also in lower LLs $n\geq1$. Note that this approximate expression is shown primarily for reasons of illustration here. In the calculations presented below, we use the exact form of the interaction potential, obtained by a Fourier transformation of Eq.\ (\ref{equ002}). Apart from small oscillations around the approximate form (black line in Fig.\,\ref{fig00}), the potential exhibits a plateau of width $2R_C$ superimposed on the bare $e^2/\epsilon r$ Coulomb potential, which is retrieved at large distances. The plateau is due to the ring-like form of the electronic wave functions in the $n$-th LL with a radius on the order of $R_C$. This model permits a common description of interacting electrons for all LLs. The Hamiltonian approach proposed by Murthy and Shankar\cite{MS} is therefore not restricted to the lowest LL, but has universal validity, provided inter-LL mixing is negligible.

For the following sections, we will choose a system of units with $\hbar\equiv l_B\equiv1$ to simplify the notation.

\section{Hartree-Fock treatment of the electron-solid phases}

Whereas the quantum nature of the electron-liquid phases, which display the FQHE, is not captured in a mean-field treatment of the model, the Hartree-Fock approximation gives reliable estimates for the energy of states with a modulated electron density such as CDWs. \cite{FKS,moessner,fogler} In this section, we follow the lines of Refs.~\cite{FKS,fogler}. The Hartree-Fock Hamiltonian turns out to be linear in the projected density operators $\bar{\rho}({\bf q})$
\begin{equation}
\label{equ004}
\hat{H}_{HF}=\frac{1}{2A}\sum_{\bf q}u_n^{HF}(q)\langle \bar{\rho}(-{\bf q})\rangle \bar{\rho}({\bf q})
\end{equation}
with $u_n^{HF}({\bf q})=v_n(\bq)-\sum_{\bf p}v_n(\bp)e^{-i(p_xq_y-p_yq_x)}/(n_BA)$. A detailed derivation of Eq.\ (\ref{equ004}) is given in the Appendix. Because the effective interaction potential in the $n$-th LL $v_n(q)$ is isotropic, the exchange potential is simply proportional to the real-space direct potential. For the intermediate LLs $n=1,2$, and $3$, the Fourier transformation may be calculated exactly without the help of the approximate scaling formula (\ref{equ0-1}), which was used in a previous study. \cite{goerbig3} A more detailed discussion of this approximation and its validity may be found in Sec.\ VI\,E. One obtains for the exchange potential in $n=1,2$, and $3$
{\small
\begin{eqnarray}
\nonumber
u_{n=1}^F(q)&=& \sqrt{\frac{\pi}{2}}\frac{e^2}{\epsilon n_B}\frac{e^{-q^2/4}}{8}\\
\nonumber
&&\times\left[\left(6-2q^2+q^4\right)I_0\left(\frac{q^2}{4}\right)-q^4 I_1\left(\frac{q^2}{4}\right)\right],\\
\nonumber
u_{n=2}^F(q)&=& \sqrt{\frac{\pi}{2}}\frac{e^2}{\epsilon n_B}\frac{e^{-q^2/4}}{128}\\
\nonumber
&&\times\left[\left(82-52q^2+44q^4-10q^6+q^8\right)I_0\left(\frac{q^2}{4}\right)\right.\\
\nonumber
&&\left.-q^4\left(30-8q^2+q^4\right) I_1\left(\frac{q^2}{4}\right)\right],\\
\nonumber
u_{n=3}^F(q)&=& \sqrt{\frac{\pi}{2}}\frac{e^2}{\epsilon n_B}\frac{e^{-q^2/4}}{4608}\left[\left(2646-2430q^2+2889q^4\right.\right.\\
\nonumber
&-&\left.\left.1236q^6+270q^8-26q^{10}+q^{12}\right)I_0\left(\frac{q^2}{4}\right)\right.\\
\nonumber
&-&\left.q^4\left(1539-828q^2+224q^4-24q^6+q^8\right) I_1\left(\frac{q^2}{4}\right)\right],
\label{equ0-2}
\end{eqnarray}}

\noindent 
with the modified Bessel functions $I_j(x)$. It can be seen from Eq.\ (\ref{equ004}) that the cohesive energy $E_{coh}=\langle \hat{H}_{HF}\rangle/\bar{N}_{el}$ is a functional of the order parameter $\Delta(\bq)=\langle \rhobar(\bq)\rangle/n_BA$
$$
E_{coh}^{CDW}(n;\nubar)=\frac{n_B}{2\nubar}\sum_{\bq}u_n^{HF}(q)|\Delta(\bq)|^2.
$$
It is the special form of the effective interaction potential in higher LLs ($n\geq1$) that is responsible for the CDW formation: because of the form factor $F_n(q)$, the effective interaction potential (\ref{equ002}) vanishes at non-zero values of the wave vector $q_0(n)$. It is therefore energetically favorable for the order parameter to have a maximum at these wave vectors, which leads to a density modulation in real space with a characteristic periodicity $\Lambda\sim 2\pi/q_0(n)$.

\subsection{Triangular CDWs - Bubble Phase}

We first investigate a CDW state of triangular symmetry, which is also called bubble phase. This phase has been proposed as a candidate for the ground state at $\nubar<1/2$ in higher LLs.\cite{FKS,moessner,fogler} Clusters of $M$ electrons form a super Wigner crystal to minimize the residual Coulomb interaction, and because these $M$-electron bubbles may be treated as classical objects, one expects the CDW to have a triangular symmetry.\cite{bonsall} The phase is described by a local guiding-center filling factor $\nubar(\br)=\sum_j \Theta(r_B-|\br-{\bf R}_j|)$, where $r_B=\sqrt{2M}$ is the radius of a $M$-electron bubble and ${\bf R}_j$ are the lattice vectors of the 2D triangular lattice. Simple geometrical considerations lead to the relation $r_B^2/\Lambda_B^2=\sqrt{3}\nubar/2\pi$ with the lattice constant $\Lambda_B$. Fourier transformation of the local filling factor yields the order parameter of the bubble phase
\begin{eqnarray}
\nonumber
\Delta_M^B(\bq)&=&\int\frac{d^2r}{A}\nubar(\br)e^{i\bq\cdot\br}\\
\nonumber
&=&\frac{2\pi\sqrt{2M}}{Aq}J_1(q\sqrt{2M})\sum_j e^{i\bq\cdot{\bf R}_j},
\end{eqnarray}
where $J_1(x)$ is the first-order Bessel function. Using 
$$\sum_{j}e^{-i\bq\cdot{\bf R}_j}=\frac{(2\pi)^2}{A_{pc}}\sum_l\delta(\bq-{\bf G}_l),
$$
with the volume of the primitive unit cell of the triangular lattice $A_{pc}=\sqrt{3}\Lambda_B^2/2$, one obtains for the cohesive energy of the $M$-electron bubble phase
\begin{equation}
\label{equ005}
E_{coh}^{B}(n;M,\nubar)=\frac{n_B\nubar}{M}\sum_l u_n^{HF}({\bf G}_l)\frac{J_1^2(\sqrt{2M}|{\bf G}_l|)}{|{\bf G}_l|^2},
\end{equation}
where ${\bf G}_l\neq 0$ are the reciprocal lattice vectors of the 2D triangular lattice.

\subsection{Unidirectional CDWs - Stripe Phase}

Around half-filling in higher LLs, a uni-directional CDW (stripe phase)
competes with the bubble phase because the latter would break the
particle-hole symmetry, which is exact at $\nubar=1/2$. Other patterns
such as a checkerboard pattern may also account for this symmetry, and
there is a theoretical prediction that stripes become unstable towards the
formation of an anisotropic Wigner crystal at $T=0$.\cite{MF} Here, 
the discussion of the CDW phases is limited to the
triangular and the uni-directional cases, which occur to be the most
relevant phases observed in experiments. 

The stripe phase is described by a local guiding-center filling factor
$\nubar(\br)=\sum_j\Theta(a/2-|x-x_j|)$, where $a$ is the width of one
stripe and $x_j=j\Lambda_S$ with integral $j$. The stripe periodicity
$\Lambda_S$ and the stripe width are obviously related to the partial
filling factor $\nubar=a/\Lambda_S$. As for the bubble phase, the order 
parameter is obtained by Fourier transformation of $\nubar(\br)$,
$$\Delta^S(\bq)=\frac{2}{L_x}\delta_{q_y,0}\frac{\sin\left(q_x\Lambda_S\nubar/2\right)}{q_x}\sum_je^{iq_xj\Lambda_S},$$
where $L_x$ is the extension of the system in the $x$-direction. This
yields the cohesive energy for the stripe phase
$$
E_{coh}^{S}(n;\Lambda_S,\nubar)=\frac{n_B}{2\pi^2\nubar}\sum_{l\neq0}
u_n^{HF}\left(q=\frac{2\pi}{\Lambda_S}l\right)\frac{\sin^2(\pi\nubar l)}{l^2}.
$$
The optimal stripe periodicity is obtained from a minimization of the
cohesive energy $\partial
E_{coh}^{S}(n;\Lambda_S,\nubar)/\partial \Lambda_S=0$. One then finds 
$\Lambda_S=2.84R_C$ for $n=1$, $\Lambda_S=2.76R_C$ for $n=2$, and $\Lambda_S=2.74R_C$ for $n=3$. In the large-$n$ limit the optimal stripe periodicity, which is essentially independent of $\nubar$, converges to the value $\Lambda_S=2.7R_C$ in agreement with previous studies by Fogler {\sl et al.}\cite{FKS}

\section{Quantum-liquid states at $\nubar=1/(2s+1)$}

The incompressible quantum-liquid states at $\nu=1/(2s+1)$ in the
lowest LL are described to very good accuracy by Laughlin's wave
functions.\cite{laughlin} A Laughlin-type state in an arbitrary LL may
be obtained from the ansatz\cite{macdonald1}
$$|\Omega_s\rangle_n=\sum_i\frac{\left(a_i^{\dagger}\right)^n}{\sqrt{n!}}|\Omega_s\rangle,$$
where $|\Omega_s\rangle$ is the Laughlin state in the lowest LL, and
$a_i^{\dagger}$ is the usual ladder operator for the $i$-th particle
connecting the different LLs. The energy of the Laughlin state in the
lowest LL is usually defined with respect to the uncorrelated liquid
and may be written as
$$
U=\frac{1}{2A}\sum_{\bq}v_0(q)[\bar{s}(q)-1],
$$
where $\bar{s}(q)=\langle\rhobar(-\bq)\rhobar(\bq)\rangle/\bar{N}_{el}$ is
the projected static structure factor. In order to obtain the energy
of the corresponding state in an arbitrary LL $n$, it is sufficient to
replace $v_0(q)$ by the effective interaction potential $v_n(q)$ from
Eq.\ (\ref{equ002}). The structure factor is the  Fourier
transform of the pair distribution function $g_s(r)=\int
z_3...z_N\langle z_1=0,z_2=r,z_3,...,z_N|\Omega_s\rangle$, which is given
by\cite{GMP,girvin}
{\small
$$g_s(r)=\left(1-e^{-r^2/2}\right)+\sum_{m=0}^{\infty}\frac{c_{2m+1}^s}{(2m+1)!}\left(\frac{r^2}{4}\right)^{2m+1}e^{-r^2/4}.$$}
The projected structure factor thus reads
$$\bar{s}(q)=(1-\nubar)+4\nubar\sum_{m=0}^{\infty}c_{2m+1}^s
L_{2m+1}(q^2)e^{-q^2/2},$$
and the energy of the Laughlin state in an arbitrary LL $n$ is
\begin{equation}
\label{equ007}
U=-\frac{\nubar}{2A}\sum_{\bq}v_n(q)+E_{coh}^L(n;s).
\end{equation}
The second term in Eq.\ (\ref{equ007}) is the cohesive energy of the Laughlin state
\begin{equation}
\label{equ008}
E_{coh}^L(n;s)=\frac{\nubar}{\pi}\sum_{m=0}^{\infty}c_{2m+1}^sV_{2m+1}^n
\end{equation}
where the interaction potential (\ref{equ002}) has been expanded in terms of 
Haldane's pseudopotentials,
$V_{2m+1}^{n}=(2\pi/A)\sum_{\bq}v_n(q)L_{2m+1}(q^2)\exp(-q^2/2)$. 
\cite{haldane1} The coefficients $c_{2m+1}^s$ may be obtained from a fit of 
the pair-distribution function of the Laughlin states to Monte-Carlo 
calculations. 
\cite{GMP,laughlin,levesque} An alternative, analytical method to obtain these
coefficients is to use the physical properties of the state described
by Laughlin's wave functions, which result in a certain number of sum
rules determining the coefficients\cite{goerbig1}
\begin{eqnarray}
\label{equ009}
\nonumber
&&\sum^{\infty}_{m=0}c_{2m+1}^s=-\frac{s}{2},\\
\nonumber
&&\sum^{\infty}_{m=0}(2m+2)c_{2m+1}^s=-\frac{s}{4},\\
\nonumber
&&\sum^{\infty}_{m=0}(2m+3)(2m+2)c_{2m+1}^s=\frac{s^2}{2},\\
\nonumber
&&c_{2m+1}^s=-1\qquad{\rm for}~m<s.
\end{eqnarray}
The first three sum rules are due to charge neutrality, perfect
screening and compressibility, respectively. The last
condition is given by the electron repulsion at short
distances, where $g(r)\sim r^{2(2s+1)}$ due to the Jastrow factors in the 
Laughlin wave function.\cite{girvin}
This method yields the coefficients
\begin{center}
\begin{tabular}{|c||c|c|c|c|c|c|c|}
\hline
~ & $c_1^s$ & $c_3^s$ & $c_5^s$ & $c_7^s$ & $c_9^s$ & $c_{11}^s$ &
$c_{13}^s$ \\ \hline \hline 
$s=1$ & -1 & 17/32 & 1/16 & -3/32 & 0  & 0 & 0 \\ \hline
$s=2$ & -1 & -1 & 7/16 & 12/8 & -13/16 & 0 & 0 \\ \hline
$s=3$ & -1 & -1 & -1 & -25/32 & 79/16 & -85/32 & 0 \\ \hline
$s=4$ & -1 & -1 & -1 & -1 & -29/8 & 47/4 & 49/8 \\ \hline
\end{tabular}
\end{center}
and $c_{2m+1}^s=0$ for larger $m$. In spite of its simplicity, this
method provides energy values for the quantum-liquid states which deviate 
less than $1\%$ from Monte-Carlo calculations. \cite{levesque} This 
accuracy is sufficient for the present comparison of the competing phases 
whose relative energy difference is on the order of $10\%$.

\section{Quasiparticle/-hole excitations at $\nubar\neq 1/(2s+1)$}

Away from the filling factors $\nubar_L=1/(2s+1)$, the finite energy of the 
excited quasiparticles (for $\nubar>\nubar_L$) and quasiholes 
(for $\nubar<\nubar_L$) must be taken into account. 
Although the quasiparticles/-holes carry an electric charge, they are 
treated in the following as non-interacting particles. This approximation 
is valid for low quasiparticles/-hole densities, {\sl i.e.} in the vicinity
of $\nubar_L=1/(2s+1)$.
One obtains for the cohesive energy of the quantum-liquid phase 
\begin{equation}
\label{equ010}
E_{coh}^{q-l}(n;s,\nubar)=E_{coh}^{L}(n;s)+[\nubar(2s+1)-1]\Delta^n(s),
\end{equation}
where $\Delta^n(s)$ is the energy of quasiparticles of charge
$1/(2s+1)$ (quasiholes of charge $-1/(2s+1)$) in units of the
electron charge. 

The quasiparticle/-hole energies may be calculated analytically within the
Hamiltonian theory proposed by Murthy and Shankar. \cite{MS} The
Hamiltonian (\ref{equ001}) is investigated in a CF basis, where the CF
consists of an electron and a vortex-like excitation of vorticity $2s$ and 
charge $-c^2=-2ps/(2ps+1)$, in units of the electron charge. If one neglects
the internal structure of the CF, its electric charge
$e^*=(1-c^2)e$ leads to a reduced coupling to the external magnetic
field. These CFs populate CF-LLs as do electrons in an external magnetic 
field. The integer $p$ determines how many CF-LLs are completely filled, and 
the FQHE at fillings $\nubar=p/(2ps+1)$ may be interpreted as an IQHE of
CFs. For the
present investigations of quasiparticles/-holes around these fillings, it is
sufficient to use the ``preferred'' combination for the density 
operators \cite{MS}  
$$\rhobar^p(\bq)=\rhobar(\bq)-c^2\chibar(\bq),$$
where $\chibar(\bq)$ is the density operator of the vortex-like
excitations (``pseudovortex''). Because of its charge $-c^2$, the pseudovortex
density satisfies similar commutation relations as the original electron
density operators (\ref{equ003}),
$$
[\chibar(\bq),\chibar(\bk)]=-2i\sin\left(\frac{(\bq\times\bk)_z}{2c^2}\right)\chibar(\bq+\bk),
$$
whereas both densities commute, $[\rhobar(\bq),\chibar(\bk)]=0$. At 
$\nubar=1/2$, which corresponds to the limit $p\rightarrow\infty$ and thus to
$c^2=-1$, one obtains the algebra proposed by Pasquier and Haldane. \cite{PH}

The particular choice of the preferred combination respects the
commutation relations (\ref{equ003}) in the small-wave-vector limit,
whereas higher order corrections at larger wave vectors are strongly
suppressed by the Gaussian in the effective interaction potential
(\ref{equ002}). In terms of CF creation and annihilation operators
$c_{j,m}^{\dagger}$ and $c_{j,m}$, respectively, the density operator
becomes\cite{MS}
$$\rhobar^p(\bq)=\sum_{j,j';m,m'}\langle m|e^{-i\bq\cdot{\bf
    R}}|m'\rangle \langle j|\rhobar^p(\bq)|j'\rangle
c_{j,m}^{\dagger}c_{j',m'}$$ 
with the matrix elements ($m\geq m'$)
\begin{eqnarray}
\nonumber
\langle m|e^{-i\bq\cdot{\bf
    R}}|m'\rangle&=&\sqrt{\frac{m'!}{m!}}\left(\frac{-i(q_x+iq_y)l_B^*}{\sqrt{2}}\right)^{m-m'}\\
\nonumber
&&\times\ L_{m'}^{m-m'}\left(\frac{q^2l_B^{*2}}{2}\right)e^{-q^2l_B^{*2}/4}
\end{eqnarray}
and ($j\geq j'$)
{\small
\begin{eqnarray}
\nonumber
\langle j|\rhobar^p(\bq)|j'\rangle=\sqrt{\frac{j'!}{j!}}\left(\frac{-i(q_x-iq_y)l_B^{*}c}{\sqrt{2}}\right)^{j-j'}e^{-q^2l_B^{*2}c^2/4}\\
\nonumber
\times\left[L_{j'}^{j-j'}\left(\frac{q^2l_B^{*2}c^2}{2}\right)-c^{2(1-j+j')}e^{-q^2/2c^2}L_{j'}^{j-j'}\left(\frac{q^2l_B^{*2}}{2c^2}\right)\right],
\end{eqnarray}}
where $l_B^*=1/\sqrt{1-c^2}$ is the CF magnetic length and $L_j^m(x)$
are associated Laguerre polynomials. The quantum number $j$ describes
the CF-LL, and $m$ is the CF guiding-center number. The ground state
of the theory at $\nubar=p/(2ps+1)$ consists of $p$ completely filled
LLs and is therefore characterized by the expectation value $\langle
c_{n,j}^{\dagger}c_{j',m'}\rangle=\delta_{j,j'}\delta_{m,m'}\Theta(p-1-j)$.
In the Hartree-Fock approximation, the quasiparticle energies are thus given by the expression
$$
\Delta_{qp}^n(s,p)=\langle c_{p,m}\Hhat_n c_{p,m}^{\dagger}\rangle - \langle \Hhat_n \rangle,
$$
and the quasihole energies are
$$
\Delta_{qh}^n(s,p)=\langle c_{p-1,m}^{\dagger}\Hhat_n c_{p-1,m}\rangle - \langle \Hhat_n \rangle,
$$
where one averages over the ground state with the help of Wick
contractions. This yields
\begin{eqnarray}
\nonumber
\label{equ011}
\Delta_{qp}^n(s,p)&=&\frac{1}{2}\int_{\bq}v_n(q)\langle p|\rhobar^p(-\bq)\rhobar^p(\bq)|p \rangle\\
&&-\int_{\bq}v_n(q)\sum_{j'=0}^{p-1}|\langle p|\rhobar^p(\bq)|j'\rangle|^2
\end{eqnarray}
and 
\begin{eqnarray}
\nonumber
\label{equ012}
\Delta_{qh}^n(s,p)&=&-\frac{1}{2}\int_{\bq}v_n(q)\langle p-1|\rhobar^p(-\bq)\rhobar^p(\bq)|p-1 \rangle\\
&&+\int_{\bq}v_n(q)\sum_{j'=0}^{p-1}|\langle p-1|\rhobar^p(\bq)|j'\rangle|^2,
\end{eqnarray}
where $\int_{\bq}=(1/A)\sum_{\bq}=\int d^2q/(2\pi)^2$.
The expressions (\ref{equ011}) and (\ref{equ012}) are generalizations
to an arbitrary LL of Murthy and Shankar's results for $n=0$.
\cite{MS} In the lowest excited LLs $n=1,2$ and $3$, one  obtains the
energies of the quasiparticle excitations for the Laughlin series
($p=1$) 
\begin{center}
\begin{tabular}{|c||c|c|c|c|}
\hline
$\Delta_{qp}^n(s,p=1)$ & $s=1$ & $s=2$ & $s=3$ & $s=4$\\ \hline \hline
$n=1$ & 0.2267 & 0.1868 & 0.1550 & 0.1316\\ \hline
$n=2$ & 0.1903 & 0.1728 & 0.1543 & 0.1376\\ \hline
$n=3$ & 0.1691 & 0.1560 & 0.1453 & 0.1342\\ \hline
\end{tabular}
\end{center}
and the energies of the quasihole excitations
\begin{center}
\begin{tabular}{|c||c|c|c|c|}
\hline
$\Delta_{qh}^n(s,p=1)$ & $s=1$ & $s=2$ & $s=3$ & $s=4$\\ \hline \hline
$n=1$ & -0.07172 & -0.07032 & -0.05887 & -0.04959\\ \hline
$n=2$ & -0.07876 & -0.07853 & -0.06728 & -0.05765\\ \hline
$n=3$ & -0.07720 & -0.07944 & -0.07004 & -0.06124\\ \hline
\end{tabular}
\end{center}
in units of $e^2/\epsilon l_B$.

\section{The different phases and the reentrance of the IQHE}

Due to quasiparticle/-hole localization, the quantum-liquid phases display the FQHE at fillings $\nubar_L=1/(2s+1)$, {\sl i.e.} a plateau in the Hall resistance at values $R_{xy}=h/e^2(N+\nubar_L)$ accompanied by a vanishing longitudinal resistance. On the other hand, the $M$-electron bubble phases are insulating because they are pinned by the underlying impurities. This suppresses the collective sliding mode, which, in the absence of a pinning potential, would result in a non-zero contribution to the electrical transport. The conduction is therefore entirely realized by the electrons in the completely filled lower LLs, and this leads to an integer-quantized Hall resistance $R_{xy}=h/e^2N$. Whereas both the Wigner crystal and the bubble phases exhibit the IQHE, the Hall resistance for the stripe phase is not quantized and retrieves its classical value.

We compare the cohesive energies of these different phases for the LLs $n=1,2$, and $3$. This allows one to determine the ground state at different partial filling factors, and the results are compared to experimental findings. The discussion is limited to the range $\nubar\leq1/2$ because the regime $\nubar\geq1/2$ is related to the former by particle-hole symmetry. The observation of this symmetry in experiments\cite{exp1,exp2,exp3} supports the validity of the approximation to treat the electrons in completely filled LLs as inert and the neglection of their spin degrees of freedom.

\subsection{The Phases in $n=1$}

\begin{figure}
\epsfysize+5.2cm
\epsffile{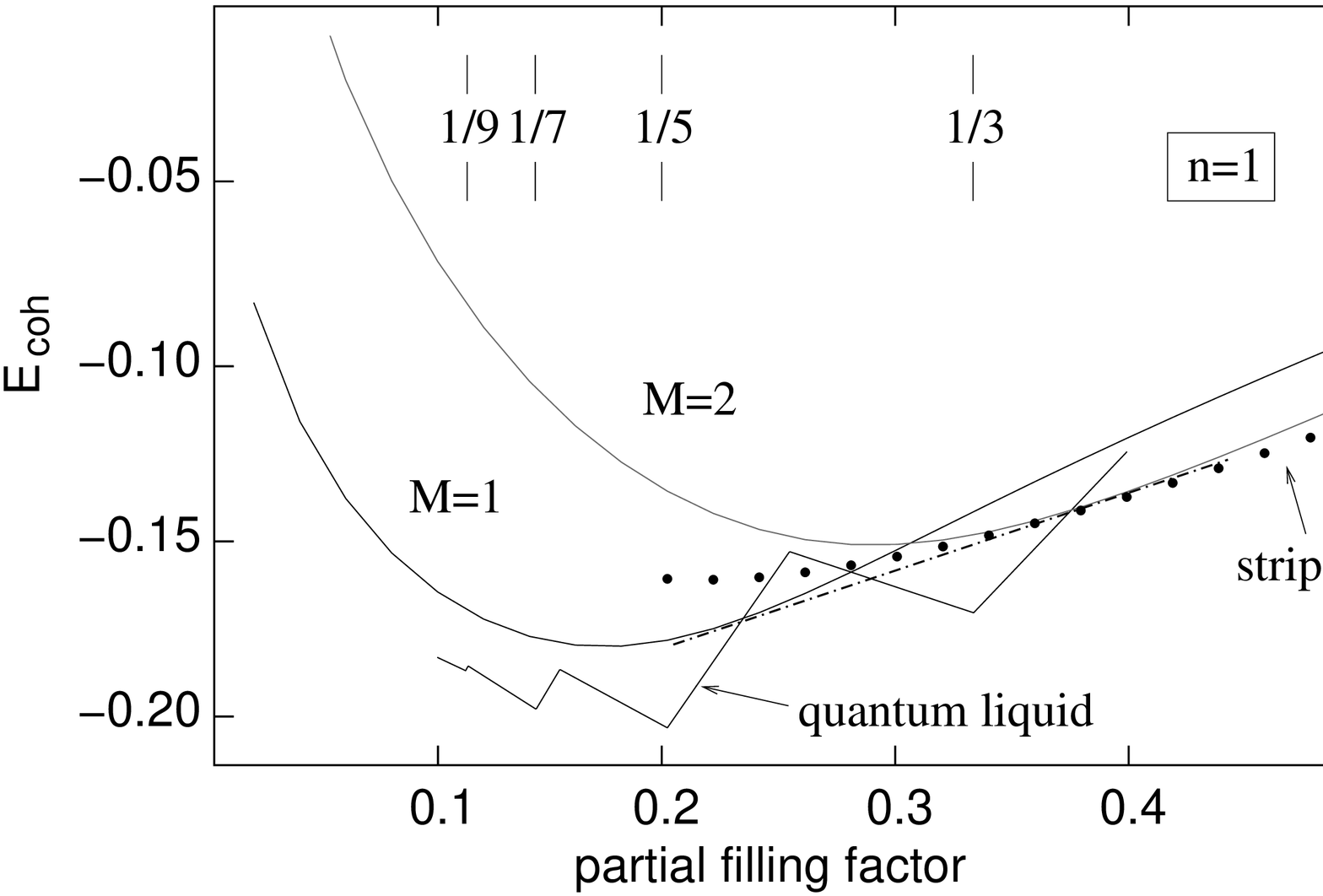}
\vspace{0.5cm}
\epsfysize+6.0cm
\epsffile{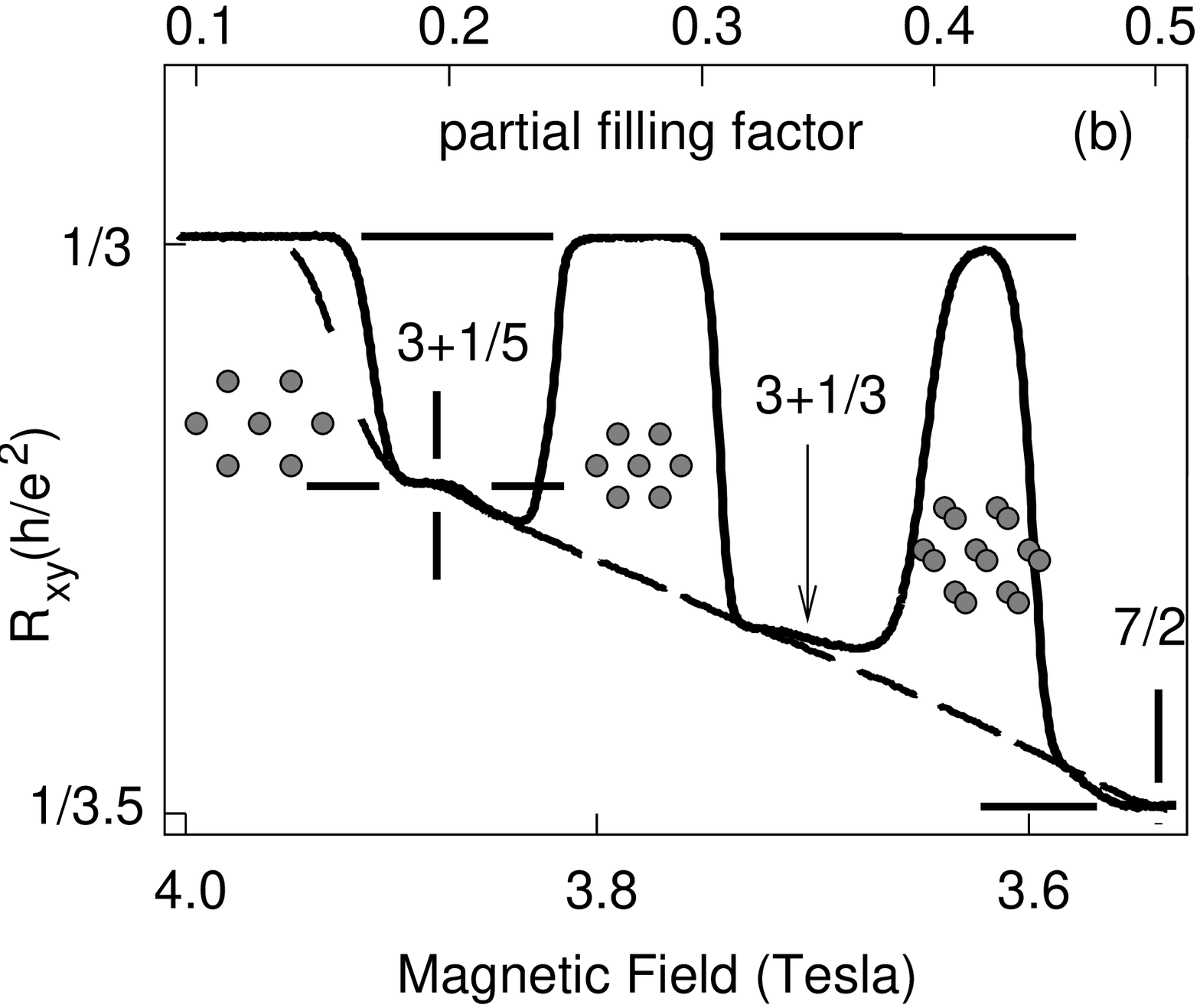}
\caption{(a) Cohesive energies of the $M$-electron 
bubble, stripe, and quantum-liquid phases for $n=1$ in units of $e^2/\epsilon l_B$.
(b) RIQHE in upper spin branch of $n=1$ measured by Eisenstein {\sl et al.}\cite{exp3} The dot-dashed tangent on the curves $M=1$ and $M=2$ shows the energy of a mixed phase of 1- and 2-electron bubbles. }
\label{fig01}
\end{figure}

Fig.\,\ref{fig01}a shows the results for the cohesive energies in $n=1$ in the absence of impurities. At low $\nubar<0.23$, the quantum-liquid phases are of a lower energy than the electron-solid phases. In the experiments by Eisenstein {\sl et al.},\cite{exp3} the FQHE is observed around $\nubar_L=1/5$, whereas they find an insulating behavior below $\nubar\lesssim0.15$ corresponding to a magnetic field above $B\sim3.9$T (Fig.\,\ref{fig01}b). A $1$-electron bubble is energetically favorable between $0.23<\nubar<0.30$. Its insulating behavior is experimentally unveiled by the RIQHE around $3.8$T. The quantum liquid has again a lower energy than the electron-solid phases between $0.30<\nubar<0.36$, and the FQHE is observed around $\nubar_L=1/3$. This alternation of electron-solid phases and FQHE states at $\nubar_L=1/3$ and $1/5$ is in agreement with recent numerical studies in the framework of the density-matrix renormalization group.\cite{shibata} The electron solid, however, appears in the numerical investigations only in form of a strong oscillation of the pair correlation function and has been identified as a stripe phase. Our energy studies show that it is the bubble phase, which is energetically favored in this range of filling factors, and that the stripe phase would be the phase of lowest energy only at higher $\nubar$. In addition, a mixed phase of $1$- and $2$-electron bubbles has a lower energy than the stripe phase up to a filling factor $\nubar\sim 0.4$ (dot-dashed tangent in Fig.\,\ref{fig01}a). This tangent represents the convex envelope of the energy curves of the $1$- and $2$-electron bubble phase. Around $\nubar=1/3$ the quantum-liquid is still energetically more favorable than such a mixed phase. The experimental observation of the RIQHE around $3.62$T supports the stability of a bubble phase at these fillings. Contrary to the presented theoretical investigations, there is experimental evidence for a stripe phase around $\nubar=1/2$ only in the presence of an in-plane magnetic field.\cite{exp4} In the absence of such an in-plane field, which breaks the rotational symmetry of the system, a quantum-liquid phase is observed in form of the even-denominator FQHE states at $\nu=5/2$ and $7/2$ in the $n=1$-LL.\cite{willet,exp3,exp4} The origin of these enigmatic states is not completely understood until now. Theoretical proposals range from spin-singlet formation\cite{HR1} to a Pfaffian wavefunction. \cite{MR1} This wavefunction, which describes a BCS-like state, may be a consequence of a pairing instability of the CF Fermi surface at $\nubar=1/2$.\cite{SPJ} A discussion of these states is, however, beyond the scope of the present article.

\subsection{The Phases in $n=2$}

\begin{figure}
\epsfysize+5.0cm
\epsffile{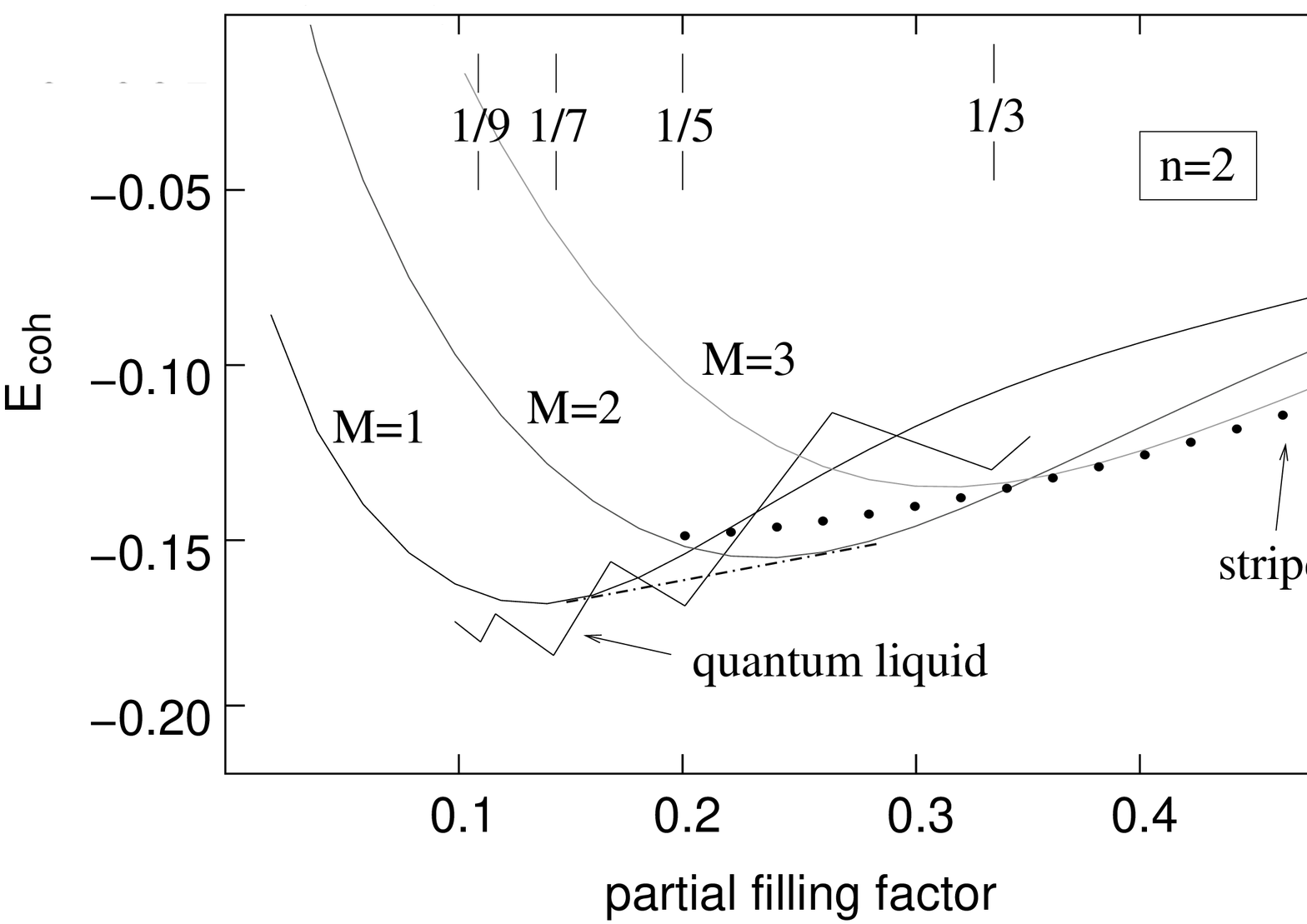}
\epsfysize+5.2cm
\hspace{0.5cm}
\epsffile{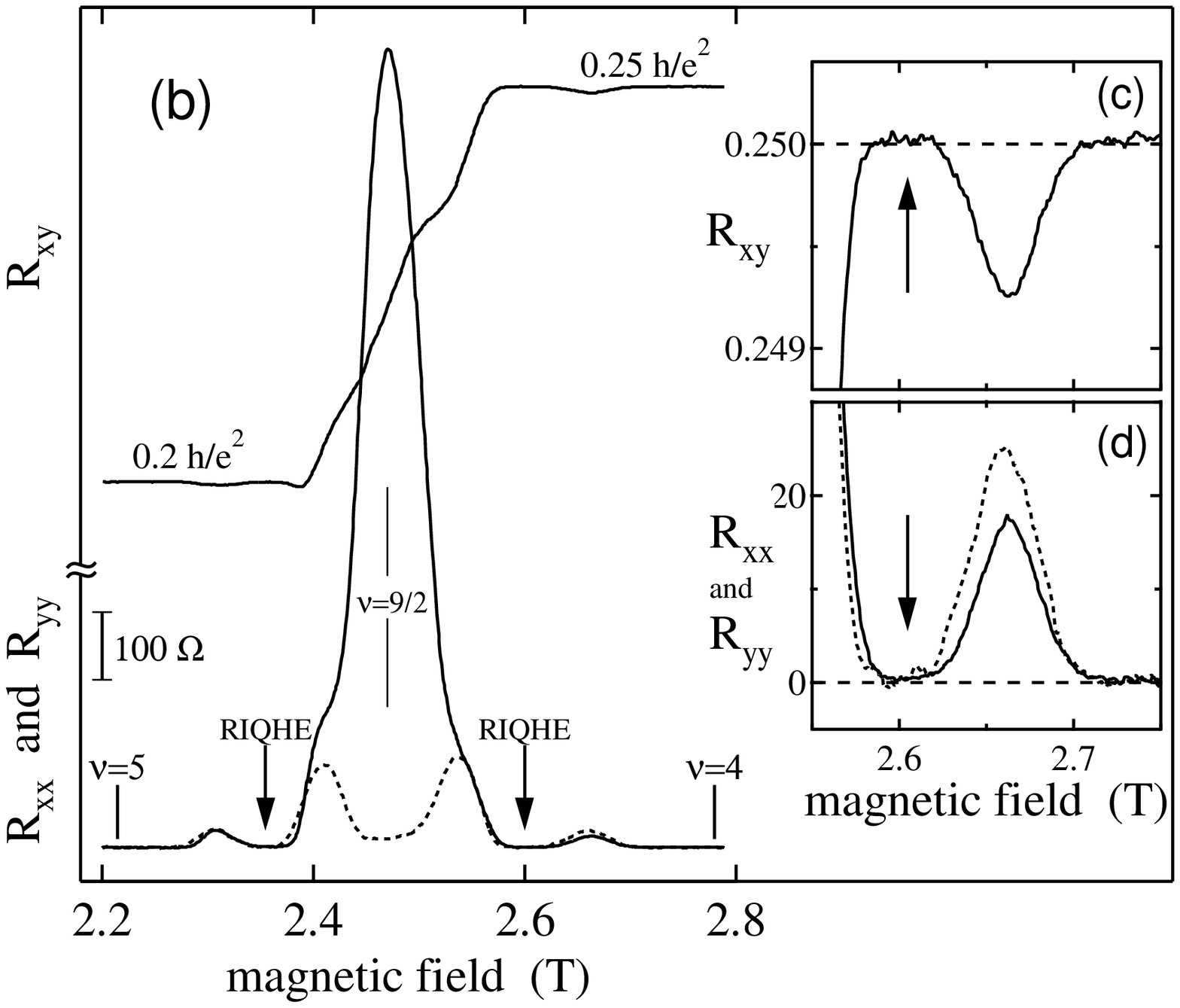}
\caption{(a) Cohesive energy for $n=2$ in units of $e^2/\epsilon l_B$.
(b) RIQHE in lower spin branch of $n=2$ observed by Cooper {\sl et al.};\cite{exp2} 
insets are a zoom on Hall (c) and longitudinal (d) resistance around $B=2.65$T.
\cite{exp2} The dot-dashed tangent on the curves $M=1$ and $M=2$ shows the energy of a mixed phase of 1- and 2-electron bubbles.}
\label{fig02}
\end{figure}

In the $n=2$-LL, the energy studies (Fig.\,\ref{fig02}a) show that the quantum-liquid phases are again favored at low densities $\nubar<0.15$ and between $0.19<\nubar<0.21$. For $0.15<\nubar<0.19$, it is the $1$-electron bubble phase which has the lowest energy. The $2$-electron bubble phase is realized in a range $0.21<\nubar<0.35$. The stripe becomes the phase of lowest energy above $\nubar\sim 0.35$, which leads to the strong anisotropy in the longitudinal magneto-resistance observed in experiments. \cite{exp1,exp2} Fig.\,\ref{fig02}b shows experimental results obtained by Cooper {\sl et al.}\cite{exp2} for the magneto-resistances in the lower spin branch of $n=2$. Apart from the anisotropy around $\nubar=1/2$ ($\nu=9/2$), a RIQHE is observed around $B=2.36$T and $2.6$T corresponding to the filling factors $\nubar=3/4$ and $1/4$. The presented energy calculations suggest that this insulating behavior is due to the formation of a $2$-electron bubble phase. The satellite maxima in the longitudinal resistance at $B\sim 2.32$T and $2.65$T, accompanied by small minima in the Hall resistance (Figs.\,\ref{fig02}c,d), indicate an incipient quantum melting towards the quantum-liquid phases around $\nubar=1/5$ or $1/7$. Whether the $1/5$ state is observable is not clear from the present energy studies because a mixed phase of $1$- and $2$-electron bubbles has approximately the same energy as the quantum liquid as is shown by the broken line in Fig.\,\ref{fig02}a.

\subsection{The Phases in $n=3$}

\begin{figure}
\epsfysize+5.5cm
\epsffile{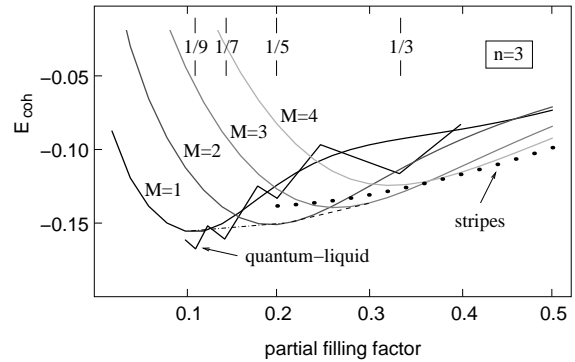}
\caption{Cohesive energy for $n=3$ in units of $e^2/\epsilon l_B$. The dot-dashed tangent shows the energy of a mixed phase of 1- and 2-electron bubbles, and the dashed tangent the energy of a mised phase of 2- and 3-electron bubbles.}
\label{fig03}
\end{figure}

In the $n=3$-LL, the quantum liquid is neither energetically favorable around $\nubar=1/3$ nor at $\nubar=1/5$ (Fig.\,\ref{fig03}). Also around $\nubar=1/7$ a mixture of $1$- and $2$-electron bubbles may have a lower energy than the quantum-liquid phase, whereas a quantum melting may be observable at $\nubar=1/9$ in extremely pure samples. The stripe phase becomes stable above $\nubar=0.36$ in agreement with experimental observations.\cite{exp1}

\subsection{General Aspects of the Phase Diagram}

With increasing $n$ the quantum-liquid phases shift to lower values of the partial filling factor, whereas the CDW phases become energetically favorable over a larger range of the electronic density in the last LL. This effect may be understood from the scaling form of the effective interaction potential in real space, as shown in Fig.\,\ref{fig00}: if the average distance $d$ between electrons in the last LL is smaller than the width $2R_C$ of the plateau, it costs only a small amount of energy to decrease the distance between two electrons below $d$. At the same time a rather large energy on the order of the height of the plateau may be gained if the electrons cluster, thus reducing the number of other electrons with which they interact strongly.\cite{FKS} The condition $d<2R_C$ leads, with the help of $d\sim1/\sqrt{\nubar}$ and $R_C=\sqrt{2n+1}$, to a critical value $\nubar_c(n)\propto 1/(2n+1)$ above which CDWs are expected to have a lower energy than the quantum-liquid phases. \cite{goerbig2} This scaling argument is supported by our energy investigations, as well as by recent numerical studies based on the time-dependent Hartree-Fock approximation. Indeed, Cot\'e {\sl et al.}\cite{cote} found that in the $n$-th LL bubble phases with up to $M=n+1$ electrons per bubble are energetically possible. Inspection of Figs.\,\ref{fig01}a, \ref{fig02}a, and \ref{fig03} confirm that as $n$ increases, bubble phases with higher $M$ are realized up to $M=n+1$. However, because large-$M$-bubble phases appear around $\nubar=1/2$, the bubble phase with $n+1$ electrons is unstable towards the formation of a stripe pattern.
 
\subsection{Comparison with the Approximate Exchange Potential}

\begin{figure}
\epsfysize+5.2cm
\epsffile{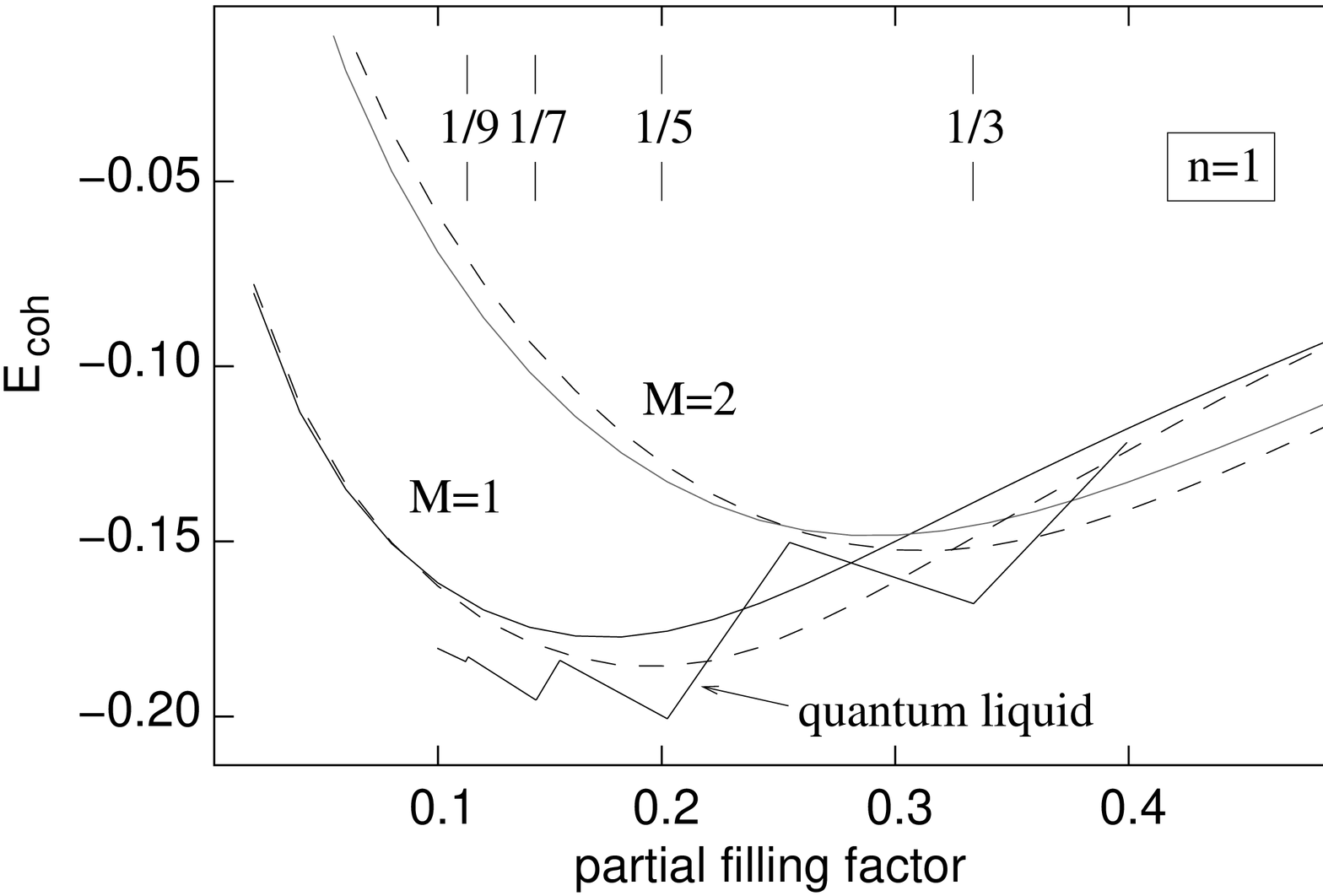}
\vspace{0.5cm}
\epsfysize+5.3cm
\epsffile{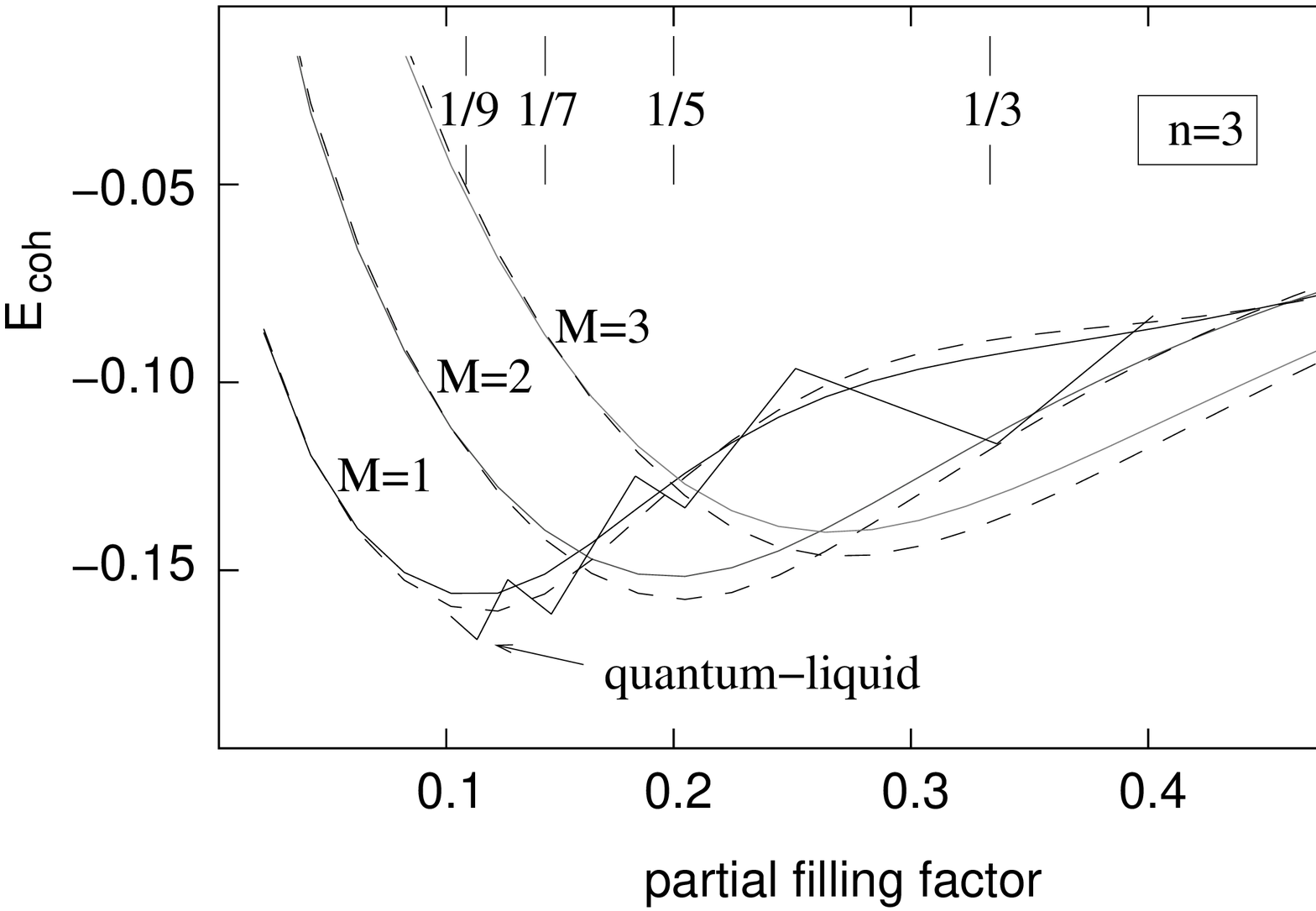}
\caption{Comparison of the results for the bubble-phase energies calculated with the help of the approximate exchange potential (dashed lines) and the results calculated with the exact exchange potential (solid lines). (a) results for $n=1$; (b) results for $n=3$.}
\label{fig05}
\end{figure}

In a recent Rapid Communication, \cite{goerbig3} we used an approximate form for the exchange potential in the calculations of the electron-solid phases. The expression 
\begin{equation}
u_n^F(q)\simeq\frac{4e^2}{\epsilon\pi^2n_B q}{\rm Re}\left[K\left(\frac{1-\sqrt{1-4(2n+1)/q^2}}{2}\right)\right]^2
\label{equ020}
\end{equation}
is derived in the same manner as the scaling form of the real-space interaction potential (\ref{equ0-1}) and becomes exact in the large-$n$ limit. For completeness, a comparison of the results in $n=1$ and $n=3$ obtained with the help of the exact and the approximate exchange potential is given in Fig.\,\ref{fig05}. As expected, the approximation yields more accurate results in $n=3$ than in $n=1$, but it captures the correct physical properties also in the lower LLs. The rather compact form of the approximate exchange potential (\ref{equ020}) may therefore be an advantage in further theoretical analyses in intermediate LLs.

\section{Competition of the Phases in the Presence of Impurities}

Although the samples used for experimental investigations of quantum Hall systems are nowadays extremely pure, the underlying residual impurities play an important role for the physical properties of these systems. On the one hand, they localize electrons/holes around $\nu=N$ (IQHE) and quasiparticles/-holes around $\nubar=p/(2ps\pm 1)$ (FQHE) and thus lead to the observed plateaus in the Hall resistances at these fillings. On the other hand, they pin the electron solids and therefore suppress their collective sliding mode. 

Whereas a weak impurity potential does not change on the average the cohesive 
energy of the quantum-liquid phases due to their incompressibility, the CDW 
phases can profit from the potential landscape by deformation of their charge 
structure. The overlap of the wave functions of electrons in different bubbles 
is negligible, and the bubbles may therefore be treated as classical objects 
with an electric charge $Me$. The elasticity constant thus approaches its 
classical value\cite{bonsall} $\mu\approx 0.25M^2e^2n_{M}^{3/2}/\epsilon$, 
where $n_M=\nubar/2\pi M$ is the bubble density.

We consider a short-range Gaussian impurity potential with correlation length
$\xi$ and strength $V_0$. In the weak-pinning limit,\cite{FL1} the energy gain
for the $M$-electron bubble phase of elasticity $\mu$ may be obtained from a 
minimization of the energy density \cite{CGL,FH} 
\begin{equation}
\label{equ013}
\varepsilon(L_0)=\frac{\mu\xi^2}{L_0^2}-V_0\frac{\sqrt{n_{el}L_0^2}}{L_0^2}
\end{equation}
with respect to $L_0$. This yields the Larkin length 
$$L_0\simeq\frac{1}{4\pi}\frac{\sqrt{M}\nubar e^2\xi^2}{\epsilon V_0},$$
and the cohesive energy of the bubble phase (\ref{equ005}) is thus 
lowered by the quantity
\begin{equation}
\label{equ014}
\delta E_{coh}^B(M,\nubar)=\frac{\varepsilon(L_0)}{n_{el}}\simeq
-\frac{(2\pi)^{3/2}}{\sqrt{M}\nubar^{3/2}}\frac{V_0^2/\xi^2}{e^2/\epsilon}.
\end{equation}
The weak-pinning limit requires that $L_0$ be larger than the lattice constant 
$\Lambda_B$, and the crossover to the strong-pinning regime may thus be 
characterized by the condition $L_0\lesssim \Lambda_B$. For the $1$-electron 
bubble phase the strong-pinning case is equivalent to mere single-particle 
localization, and the energy gain per particle is simply given by 
$\delta E_{coh}=-V_0$.

\begin{figure}
\epsfysize+5.0cm
\epsffile{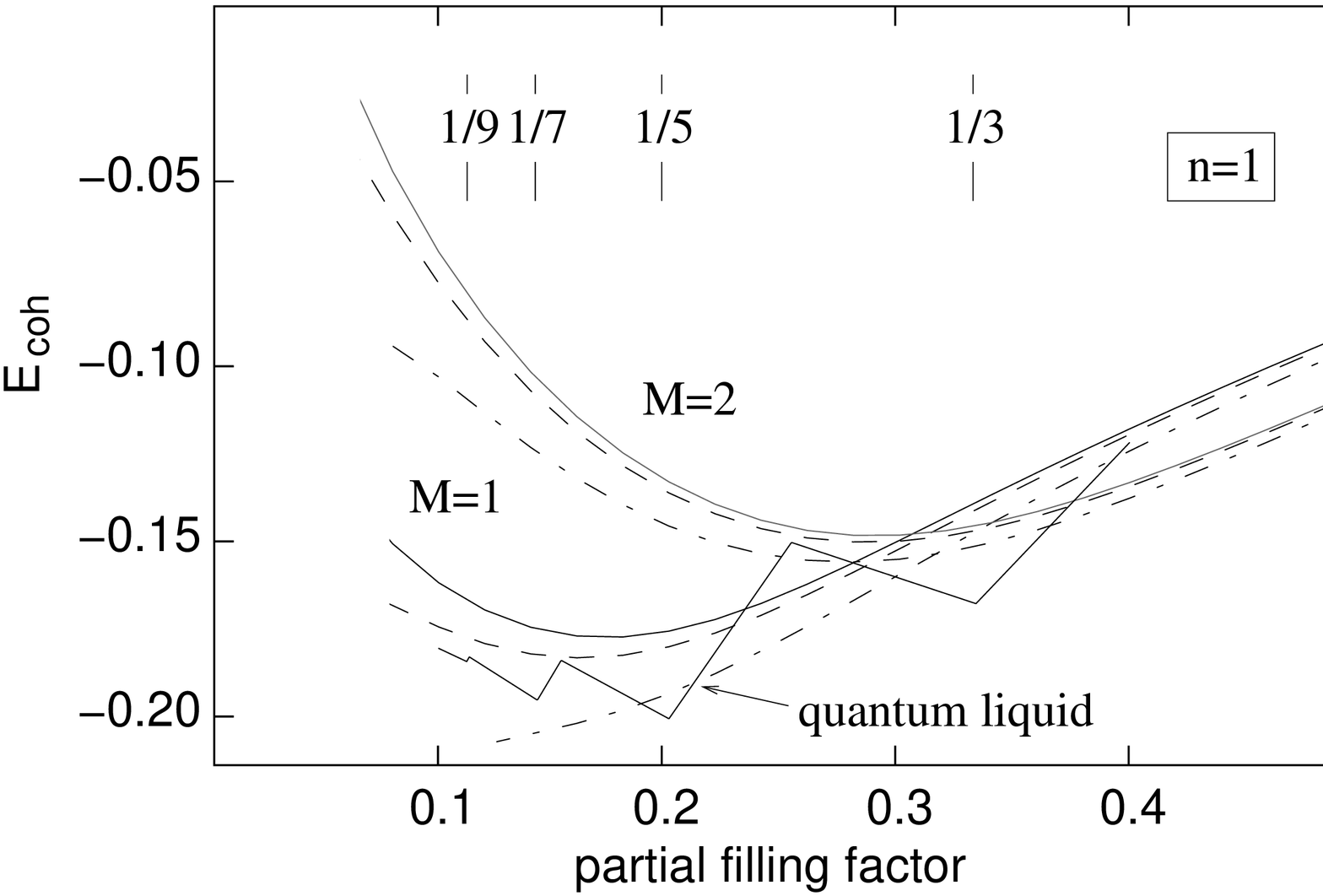}
\caption{Cohesive energies of the $1$- and $2$-electron 
bubble and quantum-liquid phases for $n=1$ in the presence of a short-range Gaussian impurity potential (in units of $e^2/\epsilon l_B$). {\sl Full line:} without impurities; {\sl dashed line:} impurity strength $V_0/\xi=0.005 e^2/\epsilon l_B^2$; {\sl dashed-dotted line:} impurity strength $V_0/\xi=0.010 e^2/\epsilon l_B^2$.}
\label{fig04}
\end{figure}

The results for the cohesive energy of the bubble phases in $n=1$ are shown in Fig.\,\ref{fig04} for two different values of the impurity strength $V_0/\xi$ in comparison to the pure case. One observes that bubble crystals with larger numbers $M$ of electrons per bubble are less affected by the impurity potential due to their higher stiffness. Furthermore, the quantum-liquid phases cease to be the energetically favored states in the dilute limit with $\nubar\rightarrow 0$, and in agreement with experimental results,\cite{chen1} we find that the Wigner crystal ($1$-electron bubble phase) is formed. The exact determination of this transition point, however, requires a detailed knowledge of the nature and the strength of the underlying impurity potential. In addition, correlation effects within the Wigner-crystal phase which modify the energy curves should be taken into account, as pointed out in previous studies for the lowest LL. \cite{lam,narevitch}

\section{Possible Experimental Investigations}

In this section, we propose some new experimental investigations, which could verify the presented theoretical results. The energy calculations indicate that the transitions between the quantum-liquid and electron-solid phases are first-order quantum phase transitions. One consequence of this result is that the phenomenon of ``super-cooling'' is expected with respect to the control parameter, which is the filling factor tuned by the magnetic field. Starting a transport measurement {\sl e.g.} from the quantum-liquid phase at $\nu=3+1/5$ and then lowering the magnetic field, the phase transition to the bubble phase would take place at a lower magnetic field than for the inverse case in which one raises the field starting from the bubble phase. The observation of such a hysteretical behavior in the Hall resistance may support the presence of first-order quantum phase transitions.

Another possible investigation of the bubble phase is via transport studies under microwave irradiation of the 2D electron system.\cite{lewis} Due to the pinning of the bubble crystal by residual impurities, the longitudinal conductivity $\Re[\sigma_{xx}(\omega)]$ depends on the microwave frequency and exhibits a resonance at a finite value $\omega_p$. The functional dependence of $\omega_p$ was recently calculated for the Wigner crystal. \cite{FH} These results may be generalized to the bubble phase if one replaces the elasticity constant $\mu$ and the Larkin length $L_0$ of the Wigner crystal by their values for the bubble phase derived in the preceding section. Depending on the strength of the magnetic field, one finds for a classical bubble crystal
$$\frac{\omega_p}{\omega_c}\sim \sqrt{\frac{\Pi_0}{m\nubar\omega_c^2}}\sim M^{-1/4}\nubar^{-3/4}, \qquad {\rm for~weak~fields}$$
and 
$$\frac{\omega_p}{\omega_c}\sim \frac{\Pi_0}{m\nubar\omega_c^2}\sim M^{-1/2}\nubar^{-3/2}, \qquad {\rm for~strong~fields},$$
where $\Pi_0\sim\mu/L_0^2$ is the characteristic self-energy of the phonons of the bubble crystal. The number of electrons per bubble may therefore in principle be determined from the resonance in the frequency-dependent conductivity at $\omega_p$. Because of the mixed phase with coexisting $M$- and $(M+1)$-bubbles, which is energetically favorable at certain filling factors, one expects to observe two peaks in $\Re[\sigma_{xx}(\omega)]$ whose respective weight varies with changing magnetic field. This effect may be easier to measure around quarter-filling of the $n=3$ than in $n=2$ because in this range of $\nu$ there is no competition with quantum-liquid or stripe phases. The resolution of such a double-peak structure, however, depends on the strength of the magnetic field. \cite{FH}

Recently, Cot\'e {\sl et al.} argued that the first-order quantum phase transitions between bubble crystals of different $M$ would lead to experimentally observable discontinuities in the magnetic susceptibility $\chi=-(1/\nubar n_B A) (\partial^2 E/\partial B^2)$ associated to the orbital magnetization. \cite{cote} It is, however, uncertain how these discontinuities will evolve if the sharp phase transition is covered by an energetically favored mixed phase of bubbles with different $M$.

\section{Conclusions}

We have presented detailed energy calculations of the competing quantum-liquid and electron-solid phases, which allow for the determination of the ground state in the intermediate LLs $n=1,2$, and $3$ as a function of the partial filling $\nubar$ of the last LL. Whereas the energies of the electron solids (triangular and uni-directional CDWs, {\sl i.e.} $M$-electron bubbles and stripes), are calculated in the Hartree-Fock approximation, which gives reliable energy estimates for states with a modulated electron density,\cite{FKS,moessner,fogler} the quantum nature of the liquid phases is not captured in a mean-field treatment of the model. 

At filling factors $\nubar_L=1/(2s+1)$, the quantum liquid may be described by Laughlin's trial wave functions. With the help of a set of sum rules imposed on these wave functions,\cite{GMP,girvin} we derived the structure factor of a quantum liquid, which has the same physical properties as the Laughlin liquid.\cite{goerbig1} The determination of the structure factor allows one to compute the energy of the quantum-liquid phase, which agrees to great accuracy with numerical results.\cite{levesque} Away from precisely $\nubar_L=1/(2s+1)$, the energy of the quantum liquid increases due to the excitation of quasiparticles/-holes of finite energy $\Delta^n(s)$, which may be calculated analytically in the Hamiltonian theory of the FQHE.\cite{MS} This non-monotonous behavior of the quantum-liquid energy gives rise to an alternation between quantum-liquid phases, which are energetically favored around $\nubar_L=1/(2s+1)$, and electron-solid phases having a lower energy in between. 

Whereas the quantum liquids display the FQHE, the electron-solid phases, which form a triangular lattice of $M$-electron bubbles, are insulating because they are pinned by underlying impurities in the sample. The alternation of the different phases results in the observed RIQHE between the FQHE states at $\nubar=1/5,1/3$, and the even-denominator state at half-filling in $n=1$.\cite{exp3} In $n=2$, this competition between the different phases is manifest in the RIQHE found around $\nubar=1/4$ and $3/4$ and in the observation of an incipient quantum melting around $\nubar\sim 1/7...1/5$.\cite{exp2} It is not clear whether the quantum melting may lead to a FQHE at $\nubar=1/5$ or $1/7$ in $n=2$. Our energy calculations indicate that the $1/7$ state is more stable than the $1/5$ state for a pure sample. Impurities, however, favor electron-solid phases preferentially at lower $\nubar$. The interpretation of the RIQHE in $n=2$ as being due to the formation of an insulating bubble phase is further supported by recent microwave experiments, which reveal a peak in the longitudinal conductivity at finite frequency.\cite{lewis} This peak is likely to be caused by the excitation of a collective depinning mode.\cite{FL1,CGL,FH} Microwave experiments in the RIQHE regime would thus clarify the origin of these insulating phases and could determine the number of electrons $M$ per bubble because of the $M$-dependence of the resonance in the frequency-dependent conductivity. However, to the knowledge of the authors, such experimental investigations remain to be performed.

\section*{Acknowledgments}

We acknowledge fruitful discussions with K.\ Borejsza, J.\ P.\ Eisenstein, T.\ Giamarchi, R.\ Moessner, V.\ Pasquier, N.\ Read, and S.\ Scheidl. This work was supported by the Swiss National Foundation for Scientific Research under grant No.~620-62868.00.

\appendix
\section{Hartree-Fock Hamiltonian}
In terms of fermionic operators, the Hamiltonian (\ref{equ001}) reads 
\begin{eqnarray}
\nonumber
\hat{H}&=&\frac{1}{2}\sum_{\bf q}v_n({\bf q})\sum_{y_0,y_0'}e^{iq_y(y_0-y_0')}\\
\nonumber
&&\times\ c_{y_0-q_x/2}^{\dagger}c_{y_0+q_x/2}c_{y_0'+q_x/2}^{\dagger}c_{y_0'-q_x/2},
\end{eqnarray}
where the indices $n$ at the operators are skipped for better legibility. This Hamiltonian becomes in the Hartree-Fock approximation \cite{FKS}
\begin{eqnarray}
\nonumber
\hat{H}_{HF}&=&\frac{1}{2}\sum_{\bf q}v_n({\bf q})\sum_{y_0,y_0'}e^{iq_y(y_0-y_0')}\\
\nonumber
&&\times\left[\langle c_{y_0-q_x/2}^{\dagger}c_{y_0+q_x/2}\rangle c_{y_0'+q_x/2}^{\dagger}c_{y_0'-q_x/2}\right.\\
\nonumber
&&\left.-\langle c_{y_0-q_x/2}^{\dagger}c_{y_0'-q_x/2}\rangle c_{y_0'+q_x/2}^{\dagger}c_{y_0+q_x/2}\right]\\
\nonumber
&=&\hat{H}_{H}-\hat{H}_{F}.
\end{eqnarray}
The Hartree term is directly given in terms of the density operator
$$\hat{H}_{H}=\frac{1}{2}\sum_{\bf q}u_H({\bf q})\langle \bar{\rho}(-{\bf q})\rangle \bar{\rho}({\bf q})$$
with the Hartree interaction potential $u_H({\bf q})=v_n({\bf q})$. The average $\langle \bar{\rho}({\bf q})\rangle$ is proportional to the order parameter of the mean-field theory. In order to write the Fock term in the same manner, we perform the variable transformations 
$$p_x=y_0-y_0', \qquad R=\frac{y_0+y_0'}{2},$$
and in a second step 
$$y_{\pm}=R\pm \frac{q_x}{2},\qquad \frac{1}{N_B}\sum_{y_+,p_y}e^{ip_y(y_+-y_--q_x)}=1,$$
where $N_B=n_BA$ is the total number of states per LL. Thus
\begin{eqnarray}
\nonumber
\hat{H}_{F}&=&\frac{1}{2}\sum_{\bf q}v_n({\bf q})\sum_{y_0,y_0'}e^{iq_y(y_0-y_0')}\\
\nonumber
&&\times\ \langle c_{y_0-q_x/2}^{\dagger}c_{y_0'-q_x/2}\rangle c_{y_0'+q_x/2}^{\dagger}c_{y_0+q_x/2}\\
\nonumber
&=&\frac{1}{2}\sum_{\bf q}v_n({\bf q})\sum_{R,p_x}e^{iq_yp_x}\langle c_{R-q_x/2+p_x/2}^{\dagger}c_{R-q_x/2-p_x/2}\rangle\\
\nonumber
&&\times\ c_{R+q_x/2-p_x/2}^{\dagger}c_{R+q_x/2+p_x/2}\\
\nonumber
&=&\frac{1}{2N_B}\sum_{\bf q}v_n({\bf q})\sum_{p_y,p_x}e^{i(q_yp_x-q_xp_y)}\sum_{y_+,y_-}e^{ip_y(y_+-y_-)}\\
\nonumber
&&\times\ \langle c_{y_-+p_x/2}^{\dagger}c_{y_--p_x/2}\rangle c_{y_+-p_x/2}^{\dagger}c_{y_++p_x/2}.
\end{eqnarray}
We finally obtain 
$$\hat{H}_{F}=\frac{1}{2}\sum_{\bf p}u_F({\bf p})\langle \bar{\rho}(-{\bf p})\rangle \bar{\rho}({\bf p}),$$
and one finds that the Fock potential is the Fourier transformed Hartree potential accompanied by an interchange of the $x$- and $y$-axes
$$
u_F({\bf p})=\frac{1}{N_B}\sum_{\bf q}v_n({\bf q})e^{-i(p_xq_y-p_yq_x)}.
$$
That both the direct and the exchange potentials are related to each other by a Fourier transformation, has already been pointed out in previous works for the case of an isotropic interaction.\cite{FKS,moessner} However, if the interaction is anisotropic, the rotation of the frame of reference by $\pi/2$ has to be taken into account in the calculation of the Fock potential. The Hartree-Fock Hamiltonian can therefore be written entirely in terms of the order parameter and the projected density operator
\begin{equation}
\label{equA01}
\hat{H}_{HF}=\frac{1}{2}\sum_{\bf q}u_{HF}({\bf q})\langle \bar{\rho}(-{\bf q})\rangle \bar{\rho}({\bf q}),
\end{equation}
with the effective Hartree-Fock potential
\begin{eqnarray}
\label{equA02}
\nonumber
u_{HF}({\bf q})&=&u_H({\bf q})-u_F({\bf q})\\
&=&v_n({\bf q})-\frac{1}{N_B}\sum_{\bf p}v_n({\bf p})e^{i(p_xq_y-p_yq_x)}.
\end{eqnarray}

\end{document}